\documentstyle[preprint,prl,aps,epsf]{revtex}
\begin{document}
\draft
\preprint{}
\title{Phase transition of a Bose gas in a harmonic potential}
\author{Kedar Damle, T. Senthil, Satya N. Majumdar and Subir Sachdev}
\address{Department of Physics, Yale University\\
P.O. Box 208120, New Haven, CT-06520-8120, USA.\\}
\date{\today}
\maketitle

\begin{abstract}
We consider a dilute Bose gas confined by a harmonic potential. We define
an appropriate thermodynamic limit and analyze the properties of the phases
and phase transition in this limit. Critical properties in the presence of the
potential are found to be different from, though simply related, to those in
the usual translationally invariant case.
We argue that the properties of
magnetically trapped rubidium\cite{Wieman} and sodium\cite{ketterle} gases (in
which Bose-Einstein condensation has been recently observed) are well
approximated by our thermodynamic limit except in a narrow window of temperature
around the critical temperature. We also consider the effect of the confining
potential on the non-equilibrium dynamics following a rapid quench to the
ordered side and give a scaling description of the late time universal
dynamics. 

\end{abstract}
\vspace{0.25cm}

\pacs{PACS numbers:03.75.Fi, 03.65.Db, 05.30.Jp, 32.80.Pj}

\narrowtext

Much excitement has been generated in the last few months by the
observation of
Bose-Einstein condensation (BEC) in magnetically trapped alkali atoms
\cite{Wieman}.
Since the original observation of BEC in rubidium atoms, two other groups
have reported evidence for the same in lithium \cite{hulet} and sodium
\cite{ketterle} atoms. A key feature of these experiments is the
presence of
a confining harmonic potential that the atoms feel. In this paper,
we study the
effects of such a confining potential on the statistical
mechanics of ideal and
interacting Bose systems (we confine our attention to repulsive
interactions -
so our results are not directly applicable to the Lithium experiment).
In particular, we study the effects of this
confining potential on the equilibrium critical properties
of the finite temperature phase transition at which the Bose condensate
first appears. We show how to define a sensible thermodynamic limit
in which the critical singularities are present; the experiments
are, of course, in finite systems - this is accounted for by
finite-size scaling crossover functions which smooth out
the singularites, in a manner quite analogous to that for familiar phase
transitions in translationally invariant systems placed in a finite box.
We will find that the exponents of the critical singularities are
related, but {\em not\/} identical, to those of translationally invariant
systems. The finite-size scaling crossover functions are expected to be new
and not simply related to known crossover functions. We also consider the 
non-equilibrium dynamics associated with the establishment of long-range order
following a rapid quench to the superfluid phase. We argue that recent results
\cite{us} 
obtained for the case of a translationally invariant system placed in a finite
box may be used to understand the effects of the confining potential on this 
dynamics.

On general grounds, there is no true
phase transition to a Bose condensed phase in the presence of a confining
potential for a finite number of particles. Such a transition is expected to
appear only in a suitable ``thermodynamic" limit; we will argue here that
the experimentally appropriate limit is one
in which the frequency, $\omega$, of the confining potential goes
to zero, and the number of particles, $N$, to infinity, while
keeping $N\omega^3$, the interaction strength, and the temperature ($T$)
fixed.
We will show that the critical properties in this limit can be
understood by an application of the ``local density approximation'' of
Oliva~\cite{Oliva}. In this approximation,
the system is viewed as a collection of
homogenous semi-macroscopic blocks, each with it's own local
chemical potential.  Each of these blocks can be treated independently
of the others, and have properties characteristic of large uniform systems.
The properties of the system with the confining potential can then be
related to the corresponding
properties of the usual translationally invariant systems. Many
aspects of our general discussion below can be checked explicitly in a 
Hartree-Fock calculation\cite{ourselves}. We thus obtain
a complete understanding of the properties of the system
in the thermodynamic limit. (In earlier treatments~\cite{Oliva,Yang},
this approach was used for an approximate treatment of the non-critical
properties of the inhomogenous Bose gases,
but the importance of the thermodynamic limit in justifying it was not noted).
For the phase transition, we estimate by
the usual Ginzburg criterion that the system crosses over to the critical
non-ideal regime only at small deviations from criticality
(${|T - T_c| / T_c} \simeq 10^{-3}$ for the existing experimental
systems); some properties of the system in this
regime are quite significantly modified by the presence of the potential.
Further, we show that the experimental systems are quite accurately described by
our thermodynamic limit except in a narrow window of temperatures
near the critical temperature, where finite-size crossovers
need to be considered. However, this window is somewhat bigger than
the temperature range for non-ideal behaviour; thus the
crossover to non-ideal critical behaviour will be complicated
by the presence of finite-size effects.

We will describe our results using the Hamiltonian (in second-quantized
notation)
\begin{equation}
{\cal H} =  \int d^3 x \left[\frac{\hbar^2}{2 m}|\nabla \psi |^2 +
          \left(\frac{1}{2}m \omega^2 (x^2 + y^2 +
            \lambda^2 z^2)
             - \mu\right)| \psi|^2 + \frac{u}{2}|\psi|^4\right]
\end{equation}
where $m$ is the mass of the bosons,
the interaction strength $u$ is related to the scattering length $a$
by $ u = 4 \pi \hbar^2 a/m$, and $\lambda$ is the anisotropy in the
harmonic potential. When $\omega$ is finite, the system is
 confined
to a finite region and will not have a true phase transition to a phase
with long-range order. We expect ``thermodynamic" behaviour to emerge only
in the limit in which $\omega$ is sent to zero and the number of
particles to infinity.
In the usual case of a system confined in a box, this is achieved
by keeping the density constant while sending the box size to infinity.
For the oscillator case, this suggests that we scale the number of
particles $N$
with the volume over which the system is
confined. This can be estimated at high temperatures (where interactions
and quantum effects may be neglected) to be
$\simeq ({k_B T / m \omega^2})^{3 \over 2}$. Thus, we guess that
the thermodynamic limit is defined by sending
$\omega$ to zero while keeping $N \omega^3$ fixed. As shown below, this does
turn out to be a perfectly sensible limit in which there is a true phase
transition at a finite non-zero temperature
and the free energy per particle is
finite.

First consider the system in its high temperature phase. As $\omega$ goes to
zero, the potential varies over a macroscopic length scale
$\sim {1 / \omega}$. Divide the system up into blocks,
such that the potential
does not vary significantly across any block. Each of these blocks is
macroscopic in size and uniform. The correlation length in any block is
some microscopic number (much smaller than the size of the block)
determined by the local chemical potential. The different blocks can therefore
be treated independently of each other. Furthermore, the properties of any block
are well approximated by the thermodynamic behaviour of the corresponding
uniform system. $\em{Extensive}$ quantities such as the free energy are then 
a sum over the free energies of the individual blocks. For the total
free energy $F$ we thus have
\begin{equation}
\label{freen1}
F = \sum_{blocks} F_{block} = \int d^3 r {\cal F}(\vec r),
\end{equation}
where ${\cal F}(\vec r)$ is the free energy density of a block of size $d^3 r$
centered at the point labelled by $\vec r$. Note that we
have replaced the sum
over blocks by an integral, as the free energy density varies
slowly from one block to another.  The only $\vec r$ and $\omega$
dependence of ${\cal F}(\vec r)$ is through the local chemical
potential $\mu (\vec r) = \mu - {1\over 2}m\omega^2(x^2 +y^2 +\lambda^2 z^2)$
and therefore ${\cal F}(\vec r)=
f(\omega x,\omega y,\omega \lambda z)$. Thus
\begin{equation}
\label{freen2}
F = {1\over \lambda \omega^3}\int d^3 \tilde{r} f(\tilde{r})
\end{equation}
where $\tilde{x} = \omega x$, $\tilde{y} = \omega y$ and $\tilde{z} = \lambda
\omega z$. Note that the function $f(\tilde{r})$ has no $\omega$ dependence.
We immediately see that the free energy per particle $F/ N$ is finite 
in the limit $N \rightarrow \infty$, $\omega \rightarrow 0$, $N \omega^3$ fixed.
Thus the properties of the high temperature phase are related trivially to the
corresponding properties of the uniform system.

This approach will fail if the correlation length in some of the blocks becomes
bigger than the block size. For a small but non-zero $\omega$, this would happen
close enough to a critical point. {\it The true thermodynamic behaviour is
accessed by first sending $\omega$ to zero and then approaching the critical
point}. When the limits are taken in this order, the block sizes can always be
taken larger than the correlation lengths and so the equations (\ref{freen1})
and (\ref{freen2}) remain valid all the way upto a critical point.
It is now easy to see
that there is indeed a phase transition at a finite non-zero value of the
temperature. As we have already seen, at high enough temperature there is a
well-defined disordered phase; now consider the system at low enough
temperature - there will always be some blocks with density bigger than the
threshold value required for condensation at that temperature. These blocks
will then become superfluid. Thus below a certain non-zero finite critical
temperature the disordered phase is unstable to superfluid ordering.

The properties of the system in the critical regime can be obtained
straightforwardly from equations like (\ref{freen1}) and (\ref{freen2}). The
singular part of the free energy density at point $\vec r$ satisfies the
hyperscaling relation $ {\cal F}(\vec r) = {C / (\xi(\vec r))^3} $,
where $\xi (\vec r)$ is the correlation length in the block centered at
$\vec r$ and $C$ is a constant. For
$\mu (\vec r)$ sufficiently close
to $\mu_c$, the critical point
of the uniform system, $\xi (\vec r) \sim (\mu_c - \mu (\vec r))^{-\nu}$ where
$\nu \simeq 0.67$ is the correlation length exponent of the three-dimensional
$XY$ universality class. The total free energy is given by
\begin{equation}
F = \int d^3 r {C \over (\xi (\vec r))^3} + \ldots
\sim (\mu_c - \mu)^{3 (\nu + {1\over 2})} + \ldots
\end{equation}
The integral is upto a value of $r$ which is of the order of, but smaller than
$\sqrt{k_B T / m \omega^2}$, and the ellipses denote terms regular
in $\mu - \mu_c$. Note the
extra factor of $3 \over 2$  in the exponent. This is a consequence of the
quadratic potential in the system and leads to a violation of naive
hyperscaling. Now $T -T_c$ is analytic in
$\mu_c - \mu$ \cite{note1} and hence the specific heat exponent is
$2 -3(\nu+{1\over 2})
\simeq -1.52$. This very weak singularity is likely to be masked by analytic
background terms even if one could access the critical region.

The order parameter correlation function for
two points separated by $\vec x$,
in a block labelled by $\vec r$, is $G(\vec r ;\vec x)
\sim \exp(-x / \xi (\vec r))/x^{1+\eta}$ for $x \gg \xi (\vec r)$.
The singular part of the order parameter susceptibility, $\chi$, (which is of
physical importance in magnetic
systems) varies as $(T - T_c)^{{3 \over{2}}-\nu (2-\eta)}$. This exponent is
{\em positive} (approximately $0.18$) implying that the susceptibility is 
{\em finite} at the transition (even though there is a divergent
correlation length associated with the transition) in contrast to the uniform
case.  

We now consider the ordered phase. This phase is characterised by a non-zero
expectation value for the Bose field $\Psi$. The magnitude of this order
parameter will be spatially inhomogenous (due to the confining
potential), while its phase will be constant.
First, we argue that there is indeed a well
defined ordered phase in the thermodynamic limit defined above. Again, we
imagine that the entire system is divided into semi-macroscopic blocks 
as before.
We expect that both the density and the magnitude of the order parameter have
short range correlations and hence their values in each block are determined by
the local chemical potential \cite{note2}.
This immediately implies that the total number of
particles scales as $1 / {\omega}^3$. Similarly, we also expect the free
energy density to be determined just by the local chemical potential,
and hence
the total free energy also scales as $1 / {\omega}^3$. Thus the free energy
per particle is again finite in the thermodynamic limit.

Since the local chemical potential is a maximum at the center of the trap,
the magnitude of the order parameter will be the largest at the center and
will decrease as one moves away from the center. It eventually becomes zero
when the local chemical potential
becomes smaller than $\mu_c$. The position of this edge, denoted by $\vec r_c$,
is given by $\mu (\vec r_c)={\mu}_c$. Near the edge, 
$|\langle \Psi \rangle| \sim
(\mu (\vec r) -\mu_c)^{\beta} \sim (r_c - r)^{\beta}$ where
$\beta \simeq 0.34$ is the usual order parameter exponent for the uniform
system. This is an important point as it may be possible to study
{\it critical} properties of the system by studying it close to the edge of
the condensate in the {\it
ordered} phase.

The fraction of particles in the $k = 0$ mode, $n(k=0)$ (which
is of direct experimental significance)
is proportional to $(\omega ^3 \int d^3 r |\langle \Psi (\vec r) \rangle |)^2$
in the thermodynamic limit.
As one approaches the critical point from the ordered phase,
the spatial extent
of the condensed region shrinks to zero as $(\mu - \mu_c)^{1 \over 2} \sim
(T_c -T)^{1 \over 2}$. The maximum value of the order parameter ({\em i.e.} at
$r = 0$)  vanishes as $(\mu -\mu_c)^{\beta} \sim (T_c - T)^{\beta}$. This
implies
that $n(k=0) \sim (T_c - T)^{2(\beta +{3 \over 2})} \sim  (T_c - T)^{3.68}$.
An explicit calculation of the order
parameter profile at $T=0$ within a Hartree-Fock approximation,
was performed in
Ref.\cite{Baym}. An extension of such a calculation to the critical regime
\cite{ourselves}
provides an illustration of the general discussion above 
(see also Ref.\cite{legget}).

As before, the free energy is determined completely once the free energy density
of the uniform system is known as a function of $\mu$. Thus the low temperature
specific heat is proportional to $T^3$, like in the uniform case, though
the prefactor will be different.

The phase of the order parameter, of course, has long range correlations.
Associated with slow variations of this phase, we have the usual sound wave.
The
character of this mode is not different from the uniform
case as long as the wavelength is small compared to a typical block size. The
sound wave velocity will be determined in the usual way
by the local superfluid
density and compressibility. This spatial variation will
cause the wave to be
``refracted" as it propogates.

The critical behaviour described above will become observable only in a window
around $T_C$ which is given by the usual Ginzburg criterion. This predicts that
the system crosses over to the critical regime
when $\xi \sim {{\lambda_T}^2 /a}$ where
$\lambda_T = ({\hbar}^2 / 2 m k_B T)^{1\over 2}$ is the thermal de
Broglie wavelength and $a$ is the scattering length. This corresponds to
$|{(T - T_c) / {T_c}}| \sim ({a /{\lambda_T     }})^2 \sim 10^{-3} - 10^{-4}$
for the Rb \cite{Wieman} and Na \cite{ketterle} experiments. Outside of this
window in the high temperature phase, the behaviour of the system can be well
approximated by the thermodynamic limit of the ideal Bose gas in a harmonic
potential and the calculations of \cite{kleppner} are expected to apply.

A small but non-zero value of $\omega$ leads to deviations from the theory
described above, which is valid in the thermodynamic limit. 
``Finite-size" corrections are
most significant near the critical point where the non-zero $\omega$
rounds off any singular behaviour of physical quantities. A crude estimate
of the width of this region where finite-size effects are important may be
obtained  as follows: Near the bottom of the well, the correlation
length varies appreciably over a length
$r \sim \sqrt{(\mu_c - \mu)/(m \omega^2 )}$. Finite-size effects will be
negligible so long as this length is much larger than $\xi(r = 0)$. In the
high temperature phase, by using the ideal bose gas expression for $\xi(r = 0)$,
this can be converted to the condition
$(T - T_c )/ T_c  < \hbar \omega / (k_B T_c ) \sim 10^{-2}$. A
similar estimate is expected to hold below $T_c$ as well. Note that this
window is at least an order of magnitude bigger than the temperature range
where the system crosses over to the critical regime. Thus ``finite-size"
effects will prohibit observation of the true critical behaviour in the
current range of experimental parameters. Increasing the scattering length
and/or decreasing the frequency of the trap will enhance the possibility
of measuring critical properties.

So far we have restricted ourselves to the equilibrium properties of the system.
However, an interesting question which may also be experimentally relevant
\cite{Wieman} is: How
does the condensate grow in time  to its final equilibrium value after a rapid
quench in temperature from above $T_c$ to below ? This non-equilibrium question
has been addressed recently \cite{us} for a translationally invariant dilute
Bose gas. It was shown that the standard phenomenology of phase-ordering
kinetics \cite{bray}(which predicts, at late times, the existence of a single
time dependent
length scale $\sim t^{1 \over {z}}$) can be used to obtain a scaling form for
the equal-time
correlation function  of the Boson field $\Psi$ :
\begin{equation}
G(r,r',t,L) \equiv \langle \psi ^* (r,t) \psi(r',t) \rangle =
|\langle \psi \rangle|^2F({|r - r'| \over {L}},{ct \over {L^z}})
\label{hyp}
\end{equation}
where $L$ is the linear size of the system, $c$ is a scale factor that depends
on the final temperature (or equivalently, the final chemical potential),
 $\langle \psi \rangle$ is the equilibrium order parameter at that
temperature and $|r- r'|$ is much
larger
than all microscopic scales in the problem. The value of $z$ was numerically 
estimated to be close to $1$ and it was argued that $z$ should be exactly equal
to $1$ \cite {us}.

We now consider the effect of the confining potential on this
dynamics. Imagine, as before, splitting the system up into many blocks (of size
$ l \sim 1 / \omega $) such that
the potential does not vary significantly within any block. As long as
$t^{1 \over {z}} \leq l$, the equal time
correlation function within each block scales as:
\begin{equation}
G(r,r',t,\omega) = |\langle \psi (R \omega) \rangle|^2
H(|r - r'| \omega,c'(R \omega)t {\omega}^z)
\label{hyp1}
\end{equation}
where $R$ is the coordinate of the centre of the block, $c'$ is a scale factor
that depends on the local chemical potential and the value of $z$ is
{\em unchanged} from the translationally invariant case. Motivated
by this we can now make the following ansatz for the scaling form valid for 
arbitrarily large values of $t$ and $|r - r'|$ 
(in particular, $r$ and $r'$ may belong to
different blocks):
\begin{equation}
G = P(r \omega, r' \omega, t {\omega}^z)
\label{hyp2}
\end{equation}
where $z$ is {\em the same as before}. Note that as $t \rightarrow \infty$ $P$
must reduce to $\langle \psi (r) \rangle \langle \psi ^{*} (r') \rangle $ which
is consistent as $\langle \psi (r) \rangle$ depends on $r$ only
through $\mu (\omega r)$.

The total number of particles in the $k=0$ mode is proportional to ${\omega}^3
\int d^3 r \int d^3 r'\, P $ and thus satisfies the scaling form:
\begin{equation}
n(k=0,t) = {\omega}^{-3}f(t \omega ^z)
\end{equation}
The fraction of particles in the $k=0$ mode is given
as ${n(k=0,t) / N} = (1 / ({N \omega ^3}))f(t\omega ^z)$. The function
$f$ of
course depends on the value of temperature we quench to and the value of the
anisotropy $\lambda$. We thus conclude that the time dependence of
$n(k=0) / N$ for fixed $T$, $N \omega ^3$ and $\lambda$ but different values
of $\omega$ {\em should exhibit scaling collapse at late times}. For small 
values of its argument $f(x) \sim x ^{{3 \over z}}$. For small enough $\omega$
(i.e for a large enough system) it may be possible to observe a corresponding
window in time
in which {\em the condensate fraction has power-law growth} ${n(k=0,t) / N}
\sim t^{{3 \over z}}$.

In summary, we have presented a rather complete understanding of the effects
of the confining potential on the equilibrium properties of a dilute Bose
gas. In particular, we have demonstrated that the critical properties of
the system in a confining potential are different from but simply related to
the uniform case. We have also provided a scaling description for the late
time universal dynamics following a rapid quench to the ordered phase.

This work was supported by NSF Grant No. DMR-92-24290 and DMR-91-20525.


\begin{references}
\bibitem{note2} For this to be true, we must require that the size of any block
is much bigger than the local Josephson correlation length.
\end{references}
\end{document}